Einstein's Discovery of Gravitational Waves 1916-1918

Galina Weinstein

12/2/16

In his 1916 ground-breaking general relativity paper Einstein had imposed a restrictive coordinate condition; his field equations were valid for a system $-g = 1$. Later, Einstein published a paper on gravitational waves. The solution presented in this paper did not satisfy the above restrictive condition. In his gravitational waves paper, Einstein concluded that gravitational fields propagate at the speed of light. The solution is the Minkowski flat metric plus a small disturbance propagating in a flat space-time. Einstein calculated the small deviation from Minkowski metric in a manner analogous to that of retarded potentials in electrodynamics. However, in obtaining the above derivation, Einstein made a mathematical error. This error caused him to obtain three different types of waves compatible with his approximate field equations: longitudinal waves, transverse waves and a new type of wave. Einstein added an Addendum in which he suggested that in a system $-g = 1$ only waves of the third type occur and these waves transport energy. He became obsessed with his system $-g = 1$. Einstein's colleagues demonstrated to him that in the coordinate system $-g = 1$ the gravitational wave of a mass point carry no energy, but Einstein tried to persuade them that he had actually not made a mistake in his gravitational waves paper. Einstein, however, eventually accepted his colleagues results and dropped the restrictive condition $-g = 1$. This finally led him to discover plane transverse gravitational waves.

Already in 1913, at the eighty-fifth congress of the German Natural Scientists and Physicists in Vienna, Einstein started to think about gravitational waves when he worked on the *Entwurf* theory. In the discussion after Einstein's 1913 Vienna talk, Max Born asked Einstein about the speed of propagation of gravitation, whether the speed would be that of the velocity of light. Einstein replied that it is extremely simple to write down the equations for the case in which the disturbance in the field is extremely small. In that case the metric tensor components differ only infinitesimally from the metric without a disturbance, and the disturbance would propagate with the velocity of light. Einstein also told Born that this is a mathematically complicated problem, because in general it is difficult to find solutions to the field equations which are non-linear. In 1916, Einstein followed these steps and studied gravitational waves.

In his 1916 review paper, "The Foundation of the General Theory of Relativity", Einstein had imposed the restrictive coordinate condition $\sqrt{-g} = 1$, and his field equations were valid for a coordinate system $\sqrt{-g} = 1$. In June 1916, Einstein presented to the Prussian Academy of Sciences a paper on gravitational waves, published later under the title, "Approximate Integration of the Field Equations of Gravitation". In this paper he considered the weak field linearized approximation. The solution presented in this paper did not satisfy the coordinate condition $\sqrt{-g} = 1$. Willem de Sitter informed Einstein that for the calculation of fields in the weak field linearized approximation it was preferable not to use this coordinate condition. Indeed, de Sitter found that Einstein's 1916 gravitational field equations could be solved for the fields emanating from the body if the coordinate condition $\sqrt{-g} = 1$ was abandoned.

In dealing with systems with small velocities compared to the velocity of light and assuming that the gravitational field is infinitely weak, Einstein solved a linearized approximation version of his field equations because this version resembled the field equations of electromagnetism. The gravitational interaction is not transmitted instantaneously. In an analogy to electrodynamics where accelerated charges emit electromagnetic waves, Einstein concluded that gravitational fields propagate at the speed of light and that plane gravity waves travel with the velocity of light $c$ in the Minkowski flat space-time (Einstein 1916b, 685).

Einstein thus first wrote the full non-linear field equations of general relativity:

$$R_{\mu\nu} - \frac{1}{2} g_{\mu\nu} R = -\kappa T_{\mu\nu}.$$

The left-hand side of the above full non-linear field equations is the Ricci tensor $R_{\mu\nu}$, constructed from the metric tensor $g_{\mu\nu}$ and its first and second derivatives.

The solution $g_{\mu\nu}$, is the Minkowski flat metric plus a small disturbance propagating in a flat space-time:

$$g_{\mu\nu} = \delta_{\mu\nu} + \gamma_{\mu\nu}.$$

$\delta_{\mu\nu}$ is the Minkowski flat metric, so that: $\delta_{\mu\nu} = 1$ if $\mu = \nu$ and $\delta_{\mu\nu} = 0$ if $\mu \neq \nu$.

Hence, the flat Minkowski metric is represented by a special relativistic system:

$$\delta_{\mu\nu} = diag(-1, -1, -1, +1),$$

and $\gamma_{\mu\nu}$ is a small deviation from the flat metric.

It was de Sitter who gave Einstein the idea to follow this route by sending him the values of the metric $g_{\mu\nu} = \delta_{\mu\nu} + \gamma_{\mu\nu}$ in a letter. Einstein confirmed that. Einstein said that de Sitter's metric led him to the simple approximate solution given in his gravitational waves 1916 paper (Einstein 1916b, 692).

Einstein introduced the $\gamma'_{\mu\nu}$ instead of the $\gamma_{\mu\nu}$ by the relations:

$$\gamma'_{\mu\nu} = \gamma_{\mu\nu} - \frac{1}{2}\delta_{\mu\nu}\gamma.$$

$$\gamma = \gamma^{\mu}_{\mu}.$$

If $\gamma_{\mu\nu}$ are everywhere very small compared to 1, we obtain a first-order approximation solution of the field equations by neglecting in the above full non-linear field equations the higher powers of $\gamma_{\mu\nu}$, and their derivatives, compared with the lower ones.

By imposing on the full non-linear field equations the harmonic coordinate condition (Einstein 1916b, 690):

$$\sum_{\nu} \frac{\partial \gamma'_{\mu\nu}}{\partial x_{\nu}} = 0,$$

one is then led to the linearized gravitational field equations:

$$\sum_{\alpha} \frac{\partial^2}{\partial x^2_{\alpha}} \gamma'_{\mu\nu} = 2\kappa T.$$

In the new harmonic coordinate system Einstein calculated $\gamma'_{\mu\nu}$ in a manner analogous to that of retarded potentials in electrodynamics (i.e., the $\gamma'_{\mu\nu}$ may be directly integrated with the help of retarded potentials):

$$\gamma'_{\mu\nu} = -\frac{\kappa}{2\pi} \int \frac{T_{\mu\nu}\left(t - \frac{r}{c}\right)}{r} dV,$$

where the expression on the right has to be taken at the time of emission, at the retarded time $t - r/c$, reaching a point at the time $t$, and $r$ is the distance to the source.

The expression:

$$\frac{\partial^2 \gamma'_{\mu\nu}}{\partial x^2_1} + \frac{\partial^2 \gamma'_{\mu\nu}}{\partial x^2_2} + \frac{\partial^2 \gamma'_{\mu\nu}}{\partial x^2_3} - \frac{1}{c^2}\frac{\partial^2 \gamma'_{\mu\nu}}{\partial x^2_4} = \left(\frac{1}{c^2}\frac{\partial^2}{\partial t^2} - \nabla^2\right)\gamma'_{\mu\nu} = 2\kappa T,$$

then represents a wave equation in Einstein's linearized gravitational field equations:

$$\sum_{\alpha} \frac{\partial^2}{\partial x^2_{\alpha}} \gamma'_{\mu\nu} = 2\kappa T.$$

Einstein needed the energy components of the gravitational field. He therefore multiplied the linearized field equations by $\frac{\partial \gamma'_{\mu\nu}}{\partial x_{\sigma}}$ and summed over μ and ν and obtained the conservation law for the total energy-momentum for matter:

$$\sum_{\nu} \frac{\partial(T_{\mu\nu} + t_{\mu\nu})}{\partial x_{\nu}} = 0,$$

where the energy components of the gravitational field $t_{\mu\nu}$ are (Einstein 1916b, 690-691):

$$t_{\mu\nu} = \frac{1}{4\kappa}\left[\sum_{\alpha\beta}\frac{\partial \gamma'_{\alpha\beta}}{\partial x_{\mu}}\frac{\partial \gamma'_{\alpha\beta}}{\partial x_{\nu}} - \frac{1}{2}\delta_{\mu\nu}\sum_{\alpha\beta\tau}\left(\frac{\partial \gamma'_{\alpha\beta}}{\partial x_{\tau}}\right)^2\right].$$

He then calculated the gravitational field of a mass point of mass *M*, resting at the origin of the coordinate system.

However, in obtaining the above derivation, Einstein made a mathematical error. In his 1918 paper "On Gravitational Waves", he explained the nature of his error (Einstein 1918, 157): He had multiplied the field equations by $\frac{\partial \gamma'_{\mu\nu}}{\partial x_{\sigma}}$ instead of by $\frac{\partial \gamma_{\mu\nu}}{\partial x_{\sigma}}$; namely, he used γ'μν instead of γμν.

This error caused Einstein to arrive at the following conclusions: he obtained three different types of gravitational waves compatible with his approximate field equations: longitudinal and transverse gravitational waves, as well as a new type of wave.

Einstein used the energy components of the gravitational field $t_{\mu\nu}$ to compute the energy transported by the longitudinal and transverse waves. He found, however, that only the third type of gravitational waves transport energy, whereas the longitudinal and transverse gravitational waves do not transport any energy. Einstein was troubled by de Sitter's suggestion and metric which seemed to him to have led to waves that transport no energy! (Einstein 1916b, 693).

To the 1916 paper on gravitational waves, delivered to the Prussian Academy of Sciences, Einstein added an Addendum and thereafter published it. In the Addendum's response, Einstein returned to unimodular coordinates, the coordinate condition $\sqrt{-g} = 1$, and found that when imposing this coordinate condition only the third type of waves carrying energy existed (and was thus real), whereas the other two types were eliminated (and were thus fictitious). He therefore assumed that $\sqrt{-g} = 1$ holds; i.e. that the coordinate condition $\sqrt{-g} = 1$ was natural.

Einstein even found a very interesting physical justification for this coordinate condition. Before publishing the Addendum Einstein explained to de Sitter why he decided to return to the coordinate condition $\sqrt{-g} = 1$. Einstein introduced two coordinate systems: a coordinate system *K*, with respect to which $\sqrt{-g} = 1$ holds everywhere (the system presented in Einstein's Addendum); and de Sitter's system denoted by *K'* (this was the system presented in Einstein's 1916 paper on gravitational waves).

Einstein was now searching for plane gravitational waves. In de Sitter's system *K'*, he found three types of gravitational waves (longitudinal, transverse, and an additional type), of which only the third type was connected to energy transportation. In Einstein's system *K*, by contrast, only this energy-carrying type was present.

Einstein explained to de Sitter that this meant that the first two types of waves, the longitudinal and transverse waves, obtained for system *K'* did not actually exist in reality; but were simulated by the coordinate system's wavelike motions with respect to a Galilean space (in the context of a coordinate system in which $\sqrt{-g} = 1$). Einstein concluded that unimodular coordinates accordingly exclude systems in which we find waves with no energy and therefore waves that do not actually exist in reality.

Subsequently, Einstein added this explanation as the Addendum to his 1916 gravitational waves paper. He explained that there was a simple way to clarify the strange result that gravitational waves which transport no energy could exist, namely that they were not real gravitational waves but rather apparent waves, fictitious waves reflecting vibrations of the reference system (i.e. de Sitter's system *K'*) in which they had been derived. If from the start one selects a coordinate system in which $\sqrt{-g} = 1$, then with this choice only the third type of waves that transport energy exist. The coordinate condition $\sqrt{-g} = 1$ eliminates gravitational waves that do not transport energy, and so these types of waves are eliminated by this choice of coordinates; and in this sense they are not "real" waves.

Einstein concluded that even though it was not preferable to restrict the choice of coordinates for the calculation of the first-order approximation, his results showed that the choice of coordinates under the restriction to a coordinate system in which $\sqrt{-g} = 1$ was physically justified (Einstein 1916b, 696).

De Sitter objected to Einstein's use of the words "real" and "apparent". To this Einstein replied that by "real" he meant a process which cannot be transformed otherwise. Einstein agreed not to use this terminology and to say that his coordinate system $\sqrt{-g} = 1$ was simple or preferable, because with this choice only waves of the third type occur (one only comes across waves that transport energy). Einstein avoided the classical terminology but remained obsessed with his coordinate system $\sqrt{-g} = 1$, to which de Sitter objected.

De Sitter then suggested to Einstein that his field equations in first-order approximation could be solved if the coordinate condition $\sqrt{-g} = 1$ was abandoned. Einstein followed this suggestion and again obtained strange results, which led him back to this coordinate condition.

Einstein, still unaware of the mathematical error he had committed in his 1916 paper (using $\gamma'_{\mu\nu}$ instead of $\gamma_{\mu\nu}$), got entangled with the coordinate condition $\sqrt{-g} = 1$ and the third type of gravitational wave he had found in 1916 (Weinstein 2015, 249-251).

In Einstein's 1916 paper a gravitational field carries energy from a mass point. In his November 1915 papers Einstein postulated that $\sqrt{-g} = 1$, and derived an expression for the energy components of the gravitational field $t_\sigma^\nu$. In his 1916 gravitational waves paper, Einstein did not assume a coordinate condition $\sqrt{-g} = 1$, and derived an expression for the energy components of the gravitational field $t_{\mu\nu}$, which enabled him to conclude that only gravitational waves of the third type carry energy (Einstein 1916b, 692-693).

In September 1917, Nordström calculated the energy of a mass point in a gravitational field. Nordström, however, used the components of the energy of the gravitational field from Einstein's 1916 general theory of relativity review paper (Einstein 1916a, 806):

$$\kappa t_\sigma^\alpha = \frac{1}{2}\delta_\sigma^\alpha g^{\mu\nu}\Gamma_{\mu\beta}^\alpha \Gamma_{\nu\alpha}^\beta - g^{\mu\nu}\Gamma_{\mu\beta}^\alpha \Gamma_{\nu\sigma}^\beta.$$

Einstein derived the above equation in a theory which applied to all systems of coordinates for which

$$\sqrt{-g} = 1.$$

Nordström found that the 44-component of the above energy of the gravitational field $t_\sigma^\nu$ of a mass point in a coordinate system $\sqrt{-g} = 1$ vanishes $t_4^4 = 0$.

Therefore, in the coordinate system $\sqrt{-g} = 1$ the gravitational field and gravitational wave of a mass point carry no energy, and this did not agree with Einstein's conclusion in the Addendum of his 1916 gravitational waves paper.

Einstein tried to persuade Nordström that he had actually not made a mistake in his 1916 paper, but in October 1917 the latter again demonstrated to Einstein that the $t_4^4$ vanishes. Nordström bluntly declared that the calculation he had made based on Einstein's formula from his 1916 review paper again yielded $t_4^4 = 0$, and therefore he could not confirm Einstein's 1916 expression for $t_\sigma^\nu$ from his gravitational waves paper.

Nordström confessed to Einstein, admitting that he had simply repeated the main calculation comprehensively and unfortunately, again obtained the same result. He told Einstein that with the best of intensions he could get no other result than that one, and he could not see where the inconsistency arose either. Nordström nevertheless spoke about the problem a colleague and they both agreed that the problem could perhaps lie in the unimodular coordinates. In 1918 Erwin Schrödinger demonstrated that under the choice of the coordinate system $\sqrt{-g} = 1$, all the $t_\sigma^\nu$ energy components of the gravitational field vanish.

Einstein, however, eventually accepted Nordström's and Schrödinger's results, and he realized that his 1916 gravitational wave paper contained calculation errors, and in 1918 he presented a cleaned-up version of the paper to the Prussian Academy. In his 1918 paper on gravitational waves Einstein dropped the result that, the coordinate system $\sqrt{-g} = 1$ was privileged because it eliminated gravitational waves that did not transport energy (Weinstein 2015, 253-256).

He reproduced de Sitter's 1916 solution, and mentioned the letter that Nordström had sent him and Schrödinger's results. Einstein explained that Nordström had already pointed out to him that if we choose the coordinate system $\sqrt{-g} = 1$, then all the energy components of the gravitational field vanish, when one calculates them to second-order by means of $t_\sigma^\nu$. Einstein therefore abandoned the unimodular coordinates and as a result he obtained two types of waves and not three, thereby resolving all the problems of his 1916 gravitational waves paper (Einstein 1918, 161).

In electrodynamics and in general in physics, waves indeed come in two types, producing motions either along (longitudinal waves) or across (transverse waves) the direction of motion. However, Einstein found that like electromagnetic waves, gravitational waves are transverse waves and that longitudinal waves were a mathematical illusion, constituted a complexity of the formulation, could be transformed away and could carry no energy.

If gravitational waves that transport no energy are not eliminated by the choice of coordinates $\sqrt{-g} = 1$, then how are they eliminated? Einstein showed that gravitational waves that transport no energy are generated by coordinate transformation from a field-free system (a flat Minkowski metric). Therefore, their existence is again only an apparent one. These solutions are still fictitious, and the only real ones are the energy carrying waves; Einstein wrote the equations of these waves travelling along the *x*-direction (Einstein 1918, 161).

In his 1918 gravitational waves paper, Einstein wrote the quadrupole formula that describes the rate of energy loss due to emission of gravitational waves from a binary mechanical system.

While correcting the 1916 paper, Einstein discovered that a source emitting gravitational waves will slowly lose energy transported away by these waves. Although in electromagnetism waves are emitted by a dipole source, in general relativity a dipole source is proscribed as it also is in gravitation, in conformity with the momentum conservation law, and therefore gravitational waves cannot be emitted by dipoles but by quadrupoles (Einstein 1918, 164). Einstein's quadrupole formula gives the result:

$$4\pi r^2 t_4^r = \frac{16}{5} I^2 \omega^6.$$

*I* represents the quadrupole moment tensor, the components of which are the components of the moment of inertia of the binary system radiating gravitational waves, $4\pi r^2 t_4^r$ is the rate of energy loss due to gravitational waves and $t_4^r$ (along the radius/radial component) is the energy component of the gravitational field. A binary system consists of two bodies at distances. The orbits are elliptical and lie on the plane. The binary system loses energy by emitting radiation, the orbital angular frequency ω increases and the distance between the two bodies diminishes.

In his 1922 paper, "The Propagation of Gravitational Waves", Eddington re-derived Einstein's 1918 quadrupole formula and obtained the following result:

$$4\pi r^2 t_4^r = \frac{32}{5} I^2 \omega^6.$$

Eddington concluded that there is a discrepancy between the two formulas, his and Einstein's, and explained that it was due to a numerical slip in one or the other investigation. It turned out that the numerical slip was in Einstein's 1918 investigation where he had accidently introduced the factor 1/2, because of a minor calculation mistake (Eddington 1922, 279).